# Enhancing Perpendicular Magnetic Anisotropy in Garnet Ferrimagnet by Interfacing with Few-Layer $WTe_2$

Guanzhong Wu[1*], Dongying Wang[1], Nishchhal Verma[1], Rahul Rao[2], Yang Cheng[1], Side Guo[1],

Guixin Cao[1], Kenji Watanabe[3], Takashi Taniguchi[4], Chun Ning Lau[1], Fengyuan Yang[1],

Mohit Randeria[1], Marc Bockrath[1], and P. Chris Hammel[1]

*1.Department of Physics, The Ohio State University, Columbus, OH, 43210, USA*

*2.Materials and Manufacturing Directorate, Air Force Research Laboratory, Wright-Patterson Air Force Base, Dayton, OH, 45433, USA*

*3. Research Center for Functional Materials, National Institute for Materials Science, 1-1 Namiki, Tsukuba 305-0044, Japan*

*4. International Center for Materials Nanoarchitectonics, National Institute for Materials Science, 1-1 Namiki, Tsukuba 305-0044, Japan*

*wu.2314@osu.edu

**Abstract:**

Engineering magnetic anisotropy in a ferro- or ferrimagnetic (FM) thin film is crucial in spintronic device. One way to modify the magnetic anisotropy is through the surface of the FM thin film. Here, we report the emergence of a perpendicular magnetic anisotropy (PMA) induced by interfacial interactions in a heterostructure comprised of a garnet ferrimagnet, $Y_3Fe_5O_{12}$ (YIG), and the low-symmetry, high spin orbit coupling (SOC) transition metal dichalcogenide, $WTe_2$. At the same time, we also observed an enhancement in Gilbert damping in the $WTe_2$ covered YIG area. Both the magnitude of interface-induced PMA and the Gilbert damping enhancement have no observable $WTe_2$ thickness dependence down to single quadruple-layer, indicating that the interfacial interaction plays a critical role. The ability of $WTe_2$ to enhance the PMA in FM thin film, combined with its previously reported capability to generate out-of-plane damping like spin torque, makes it desirable for magnetic memory applications.

Perpendicular magnetic anisotropy (PMA) in a ferromagnetic thin film is of great interest in spintronics research and applications. Ferromagnetic nano-elements with PMA overcome their shape anisotropy, greatly ease the memory cell size reduction and improves memory retention. These exceptional properties, improving the performance of magnetic devices, make PMA highly desirable for magnetic memory applications. PMA becomes even more important in the recent development of solid state magnetic random-access memory (MRAM) since it allows MRAM to have lower switching current and faster switching speed compared to in-plane magnetized materials [1,2].

Magnetic storage devices generally rely on metallic magnetic materials due to their robust electrical response. Interfacial magnetic anisotropy plays a critical role in generating PMA in metallic ferromagnets. When interfacing with a nonmagnetic material (NM), electron orbital angular momentum of the magnetic ions at the ferromagnet surface will be modified, in some cases enabling strong covalent bonding, resulting in distinct magnetic properties compared to the single layer [3-6]. However, spintronics devices made of metallic magnetic materials are inherently energy consumptive due to resistive losses. Recently, complex oxide ferro- or ferrimagnet insulators (FMI) have attracted substantial interest due to their ability to transport spin excitations with low dissipation [7]. Inducing PMA in FMIs naturally becomes an important topic both for scientific and technological reasons. Several successful routes to achieving PMA in FMIs have been reported using bulk intrinsic anisotropy [8] or lattice strain [9-12]. But in most experiments, the sign of the resulting interfacial anisotropy in FMI/NM heterostructures is such as to enhance the easy-plane anisotropy [13-15]. Only one recent experiment has shown the possibility of generating interfacial PMA, and this was attributed to topological surface states [16]. Nevertheless, these results demonstrate the possibility of controlling magnetic anisotropy through interfacial interactions in FMI/NM heterostructures. Here, we report a study on YIG/WTe$_2$/hBN heterostructures, which shows that when interfacing with a low symmetry nonmagnetic van der Waals material, WTe$_2$, an additional interface-induced PMA (iPMA) term emerges in the magnetic anisotropy of the YIG thin film. The absence of topological surface states at room temperature in WTe$_2$ [17, 18] forces us to seek an explanation for our observation of enhanced PMA that is distinct from that proposed for topological insulator/YIG bilayers [16]. We therefore turn to an analysis of the broken symmetries in WTe$_2$. We point out that low symmetry WTe$_2$ has recently shown the capability of generating both in-plane and out-of-plane spin polarization in charge-spin conversion experiments [19-22]. It also enables field-free switching of PMA magnetic material, which eases the application of PMA materials in MRAM applications [23-25].

Ferrimagnetic insulator YIG is of significant research interest in spintronics due to its exceptionally low Gilbert damping [26], which describes the relaxation rate of magnetization precession. And 1T'-WTe$_2$ is a semi-metallic transition metal dichalcogenide (TMD) layered material with strong SOC [27, 28]. The crystal structure of 1T'-WTe$_2$ lacks twofold rotational symmetry about the c-axis (Fig. 1a). The only symmetry in the WTe$_2$ crystal lattice *ab* plane is the mirror symmetry about the *bc* plane [29]. This unique symmetry breaking allows out-of-plane damping-like torque to be generated [30, 31], enabling efficient switching of the out-of-plane magnetization of the adjacent magnetic material [24].

A 20nm thick YIG thin film used in our experiment is epitaxially grown on (111)-oriented Gd$_3$Ga$_5$O$_{12}$ (GGG) substrate by off-axis sputtering [32]. WTe$_2$ flakes are then mechanically exfoliated from a flux-grown crystal, and dry transferred onto the clean top YIG surface without touching any other substances. This whole process is carried out in an Ar-filled glove box with <0.1 ppm of H$_2$O and O$_2$ to protect the flakes from degradation and ensure the cleanliness of the YIG/WTe$_2$ interface. We employ hexagonal boron nitride (hBN) encapsulation to protect the WTe$_2$ flakes from oxidation after being removed from the glove box. We make two samples and focus on the data taken from sample 1 in the main text. The raw data taken from sample 2 can be found in Supporting Information Fig. S2.

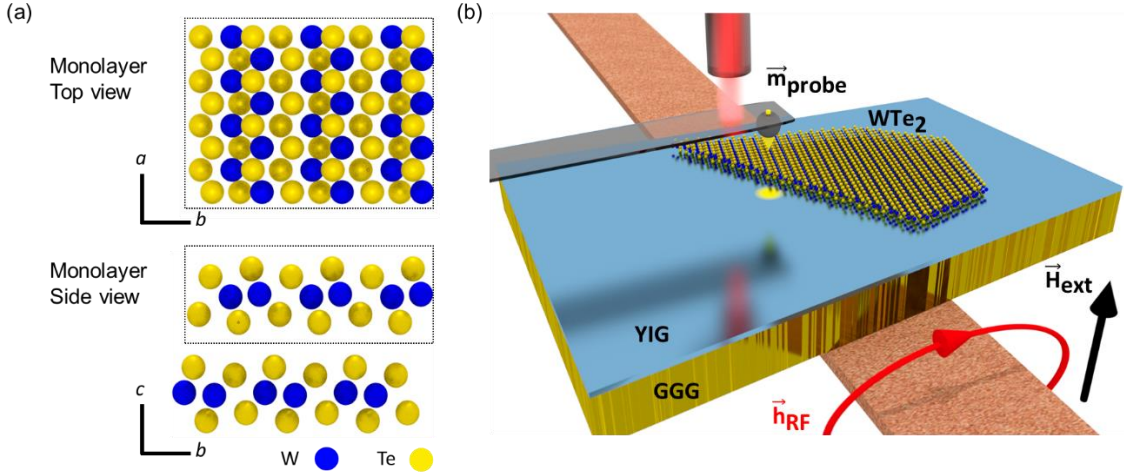

**Fig. 1 Crystal structure of WTe$_2$ and sample schematic**. **a)** Crystal lattice structure of WTe$_2$ viewed from the top along the *c*-axis and looking from the side along the *a*-axis. The black dashed box in the side view indicates a monolayer of WTe$_2$. **b)** Schematic of the ferromagnetic resonance force microscope. RF excitation is generated by a stripline underneath the sample, where the hBN encapsulation is not shown. The region of localized mode is shown as a yellow dot adjacent to the WTe$_2$ flake, and the probe magnetic moment is shown as a yellow arrow on the particle. The cantilever oscillation is detected by a fiber laser interferometer.

Figure 2a shows an optical image of the sample 1. Due to the small lateral size of the exfoliated WTe$_2$ and hBN flakes having length scales of 10 µm, we use a home-built ferromagnetic resonance force microscope (FMRFM) to measure the local ferromagnetic resonance (FMR) signal. FMRFM is a sensitive technique to detect the local magnetic properties with high spatial and spectral resolution [33]. In our FMRFM, the external magnetic field $\vec{H}_{\text{ext}}$ is aligned perpendicular to the sample plane. The cantilever tip holds a high coercivity SmCo$_5$ magnetic particle, whose moment is magnetized in the direction opposite to $\vec{H}_{\text{ext}}$ to create a magnetic field well. The field well supports a set of localized standing spin wave modes (LMs). During the measurement, we excite spin precession uniformly by a stripline underneath the sample at a fixed RF frequency (2 GHz) and sweep the magnetic field. The resonance of each LM generates a stray field, which can then be detected by the $SmCo_5$ magnetic particle attached on the cantilever through their magnetic dipole-dipole interaction (Fig. 1b). During the measurement, we keep the probe-to-sample separation around 4 µm. The operation of FMRFM is described in detail in Refs. [34-36]. For reference, we separate a region of YIG that does not contain $WTe_2$/hBN heterostructures and measure its Gilbert damping using broadband FMR. To eliminate two-magnon scattering, we perform broadband FMR in the out-of-plane field geometry. The FMR linewidth as a function of frequency measured on bare YIG (sample 1) shows a linear dependence (Fig. 2b), from which we can extract the Gilbert damping of bare YIG $\alpha_{\text{YIG}} = 1.05 \times 10^{-3}$. We also confirm that the $WTe_2$ used in the experiment is indeed the 1T' phase through polarized Raman measurements. The polarization angle dependence of the Raman peak at 212 cm$^{-1}$ (spectrum is shown in Fig. S4) exhibits minimum intensity when the excitation laser polarization is along the crystallographic $a$ axis of $WTe_2$ [37] as shown by the polar plot in Fig. 2c and Raman intensity plot in Fig. 2d.

We find the position of the YIG/$WTe_2$/hBN heterostructure with the assistance of magnetic alignment markers (Fig. 2a). Figure 2e shows two raw FMRFM scans taken in the region of YIG/hBN and YIG/$WTe_2$/hBN, indicated by the blue and the red dot in Fig. 2a, respectively, which reveals the change in FMRFM spectra at two different locations. Here we focus on the $n = 1$ LM because it has the mode radius of around 1 µm and gives the highest spatial resolution. Higher order modes have increasing mode radius and therefore, detect less localized magnetic properties. This is the reason why the quasi-uniform mode at ~ 3325 Oe does not show obvious change in resonance field or signal amplitude. We further take a line scan across the edge of WTe$_2$ flake (Fig. 2f) to resolve the spatial evolution of FMRFM spectra. The line scan in Fig. 2f (along the dashed line shown in Fig. 2a) shows three main features: first, the magnitude of the LM resonance signal is reduced in the YIG/WTe$_2$/hBN region compared to the YIG/hBN region; second, the LM resonance field for all LMs is decreased by ~40 Oe in the YIG/WTe$_2$/hBN region; third, the LMs show complex splitting and crossing when the probe is close to the boundary ($-5\ \mu m < X < 10\ \mu m$).

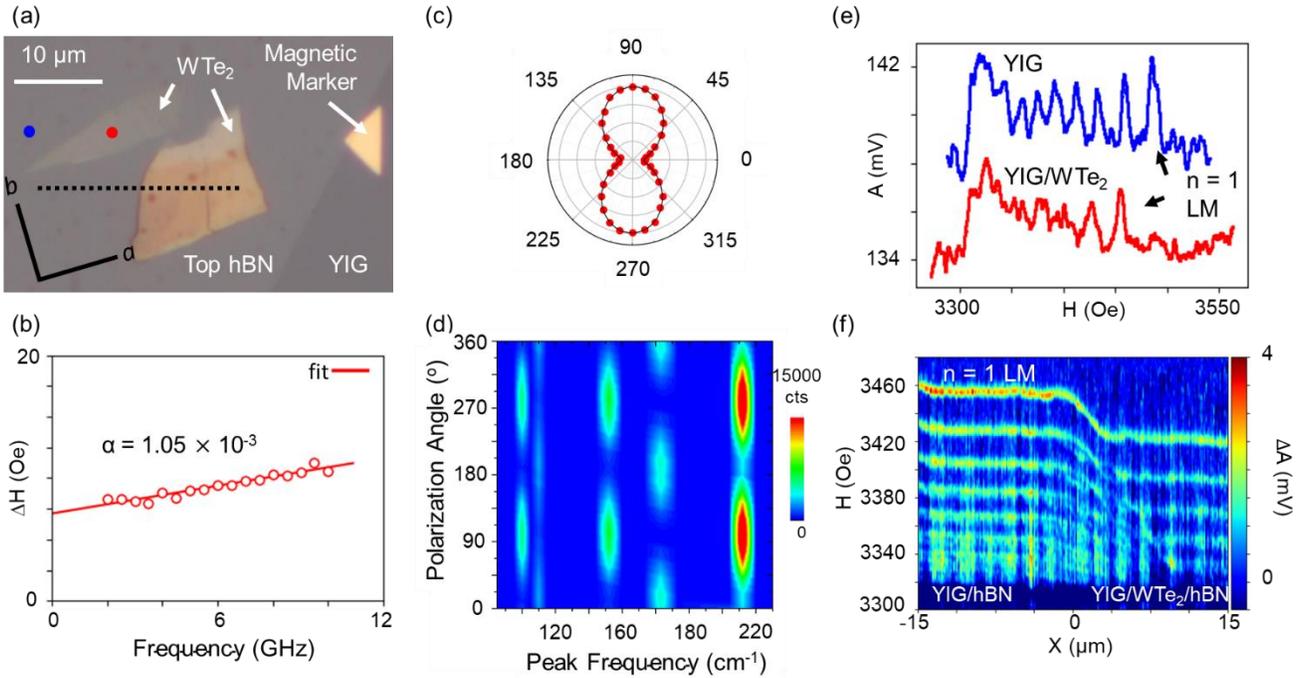

**Fig. 2 FMRFM and Raman measurement data. a)** An optical micrograph of the YIG/WTe$_2$/hBN heterostructure under study. WTe$_2$ crystal *a* and *b* axis are labeled. **b)** Broadband FMR measurement of the frequency-dependent linewidth of the YIG thin film. The measurement is done on the same piece of YIG used to make sample shown in Fig. 1b. **c)** Polar plot of the 212 cm$^{-1}$ peak Raman intensity. Angle denotes the relative angle between the measurement laser polarization and the WTe$_2$ *a* axis. **d)** 2D intensity plot showing Raman peak intensities versus polarization angle. **e)** FMRFM spectra, one over the YIG/hBN region (blue line) and the second over the YIG/WTe$_2$/hBN region (red line); these locations are indicated by the blue and red dots in Fig. 2a respectively. **f)** Color plot of field-dependence

FMRFM scans as a function of position along the trace indicated by the black dashed line in Fig. 2a. A constant background is subtracted to show only the signal from the several LM resonances.

In the following, we will explain the origin of the three observed effects using spin pumping and magnetic anisotropy. The first effect, *i.e.* signal reduction in the YIG/WTe$_2$/hBN area relative to the YIG/hBN area, is the result of enhanced relaxation due to spin pumping from YIG to WTe$_2$ [38]. The $n = 1$ LM resonance signal amplitude $\Delta A$ is inversely proportional to the square of Gilbert damping, $\alpha^2$. We determine the Gilbert damping constant $\alpha$ for YIG/WTe$_2$/hBN using $\alpha_{\text{YIG/WTe}_2/\text{hBN}} = \alpha_{\text{YIG/hBN}} \times \sqrt{\Delta A_{\text{YIG/hBN}}/\Delta A_{\text{YIG/WTe}_2/\text{hBN}}}$ (see Ref. [39]), where $\alpha_{\text{YIG/hBN}}$ is assumed to be the same as $\alpha_{\text{YIG}} = 1.05 \times 10^{-3}$ due to the low SOC and insulating character of hBN. The second effect is the decrease of $n = 1$ LM resonance field $H_{r,1}$ by ~40 Oe. And the third effect is splitting and crossing of complex modes in the region $-5$ μm $< X <$ 10 μm. The second and the third effects are due to an abrupt change of uniaxial anisotropy across the boundary separating the YIG/WTe$_2$/hBN and YIG/hBN regions [15]. Here, the uniaxial anisotropy refers to the magnetic free energy depends on the angle between magnetization and sample normal $\mathcal{F}_\text{u} = -K_\text{u}\boldsymbol{m}_\text{z}^2$, where $\boldsymbol{m}_\text{z}$ is the component of magnetization unit vector in the direction normal to sample plane and $K_\text{u}$ is the uniaxial anisotropy constant specific to sample and depends on the total interaction in the sample. When $K_\text{u}$ is positive, $\mathcal{F}_\text{u}$ is called to be of PMA type, on the other hand, if $K_\text{u}$ is negative, $\mathcal{F}_\text{u}$ is called to be of easy-plane type. This uniaxial anisotropy will lead to an effective uniaxial magnetic field $\boldsymbol{H}_\text{u} = -\partial\mathcal{F}_\text{u}/\partial\boldsymbol{M}$, where $\boldsymbol{M}$ is the magnetization. And therefore, a change in $K_\text{u}$ can modify the resonance field in a FMR measurement. In FMRFM spatial mapping, an abrupt change in $K_\text{u}$ spatially could disturb the LM and lead to mode splitting and crossing as described in Ref. 15. Moreover, in striking contrast to the previously studied

YIG/Au interface [15], which results in a 32 Oe increase of $H_{r,1}$ due to the enhanced easy-plane anisotropy, the observed decrease of $H_{r,1}$ indicates that the WTe$_2$ overlayer induces an iPMA in YIG. We note that the magnitude of the shift in $H_{r,1}$ is comparable to the easy-plane anisotropy induced by a heavy metal [15, 40] or the iPMA generated by topological surface state [16] on garnet ferrimagnetic material.

In order to probe the global effect of a WTe$_2$ overlayer on YIG, we spatially map $H_{r,1}$ using the $n = 1$ LM. Figure 3a presents an optical image of WTe$_2$ flakes on a Si/SiO$_2$ (285nm) substrate, where different colors of WTe$_2$ flakes indicate different WTe$_2$ thicknesses. Figure 3b and 3c show spatial maps of magnetic properties in the region enclosed by the black dashed rectangle in Fig. 3a. We acquire the maps using the procedure described in Ref. 39, *i.e.*, simultaneously measuring spatial variation of the magnetic anisotropy and Gilbert damping using the $n = 1$ LM resonance field $H_{r,1}$ and signal amplitude $\Delta A$. The entire WTe$_2$-covered area shows uniformly lowered $H_{r,1}$ and increased Gilbert damping relative to the area without WTe$_2$. In Fig. 3c, despite the not great signal to noise ratio in damping imaging, there is a clear Gilbert damping enhancement in WTe$_2$-covered area. The averaged Gilbert damping of YIG in WTe$_2$-covered area is $\bar{\alpha}_{\text{YIG/WTe}_2/\text{hBN}} \approx 1.30 \times 10^{-3}$, about 24% higher than $\alpha_{\text{YIG}}$. We note that due to the slight relative tilting of the scan plane and the sample plane, there is a color shift in Fig. 3b that might conceal the contrast difference in different WTe$_2$ thickness region. Therefore, to study the WTe$_2$ thickness dependence, we will show fine line scans across edges of flakes having different WTe$_2$ thicknesses.

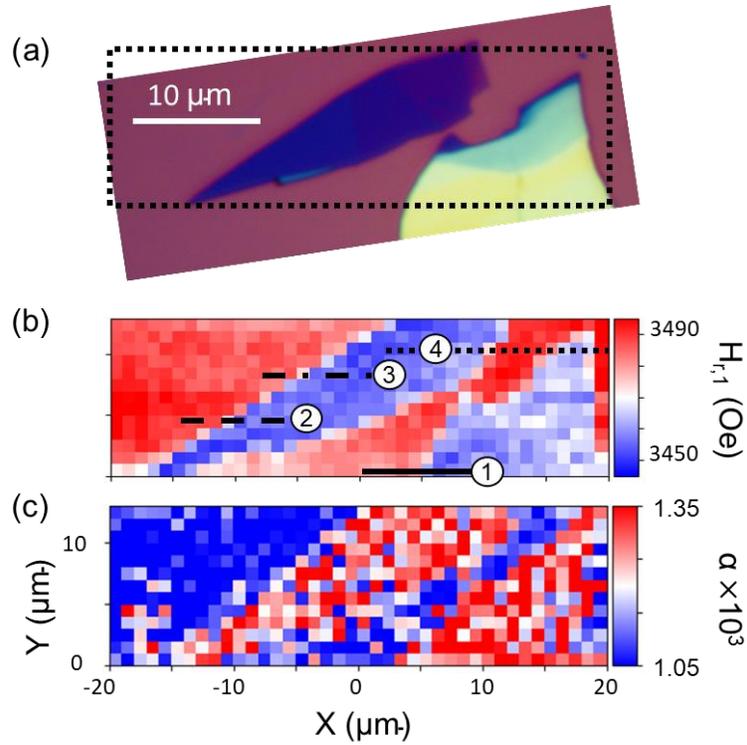

**Fig. 3 Two-dimensional FMRFM scan resolving the spatial variation of magnetic anisotropy and Gilbert damping. a)** Optical micrograph showing the color contrast of different thickness WTe$_2$ flakes (ranging from 4.7 nm to 44.8 nm) on Si/SiO$_2$(300 nm). Black dashed box outlines the FMRFM scanned area for 2D mapping. **b)** 2D map of the $n = 1$ LM resonance field. The dashed lines labeled 1-4 correspond to the four line-scans shown in Fig. S1a-S1d. **c)** 2D mapping of the Gilbert damping extracted from the $n = 1$ LM resonance peak amplitude.

Next, we want to understand what gives rise to the PMA in WTe$_2$/YIG. We rule out the effect induced by a modification of the gyromagnetic ratio by showing the resonance field shift across the WTe$_2$ edge does not depend on RF excitation frequency (See Fig. S3). We also exclude a strain induced effect given the absence of an epitaxial relation and the weakness of the van der Waals interaction between YIG and WTe$_2$. We further note that we can ignore the role of topological surface states [16] in our analysis; they are not relevant for our room temperature experiment since WTe$_2$ is a topological Weyl semimetal only below 100 K [17, 18].

We show how an analysis based on symmetry and the nature of the interfacial SOC, generalizing the theory in Ref. 41, gives insight into the PMA observed in our experiment. This will also help us understand why the *easy-axis* anisotropy we observe in WTe$_2$/YIG is so different from the results of Ref. 13, 15 on YIG interfaces with a dozen different metallic and semiconducting materials, all of which exhibit interface-induced *easy-plane* anisotropy, as is predicted by theory [41].

YIG is a ferrimagnetic Mott insulator, with two inequivalent Fe-sites coupled via antiferromagnetic (AFM) superexchange interactions. We focus on how interfacial SOC impacts AFM superexchange in YIG and show that it leads to a very specific form of the magnetic anisotropy that is governed by the direction of the effective **B**-field (see Supporting Information for a details).

Before turning to WTe$_2$/YIG, it is useful to first consider the simpler case when the only broken symmetry is the mirror plane defined by the interface. The abrupt change in lattice potential then results in an effective electric field that points normal to the interface, which in turn leads to an effective magnetic field in the rest frame of the electron that couples to its spin. Since the **E**-field points normal to the interfacial plane in which the electron moves, the resulting **B**-field arising from SOC lies *within* the interfacial plane. As we show in the SI, this leads to a SOC-induced correction to AFM superexchange that necessarily leads to an *easy-plane* anisotropy.

In the case of WTe$_2$/YIG, however, when there are additional broken symmetries. Not only does the interface break inversion symmetry, but the crystal structure of WTe$_2$ itself breaks in-plane inversion symmetry. The electric field is now no longer normal to the interface, and the effective **B**-field arising from SOC necessarily has an out-of-plane component, as shown in Fig S5b in SI. Thus, we see why the lower symmetry of WTe$_2$/YIG can naturally result in an *easy-axis* or perpendicular magnetic anisotropy (PMA); see Supporting Information for details.

We note that the lack of two-fold rotational symmetry in the *ab* plane in WTe$_2$ that plays a critical role in our understanding of PMA in WTe$_2$/YIG, has also been pointed out be crucial for the out-of-plane damping-like torque in WTe$_2$/Permalloy[30]. We note, however, that the out-of-plane damping-like torque necessarily involves current flow in WTe$_2$, while the PMA is an equilibrium property of the system independent of current flow.

We further demonstrate the interfacial origin of the observed effect by studying the influence of WTe$_2$ thickness. We show four line-scans, labeled in Fig. 3b, across the edges of WTe$_2$ with different thicknesses, ranging from 4.7 nm to 44.8 nm. From these four line-scans, we extract the $n = 1$ LM resonance field $H_{r,1}$ and the $n = 1$ LM resonance signal amplitude $\Delta A$. Figures S1a-d in the Supporting Information show the evolution of $H_{r,1}$ and $\Delta A$ along the traces labeled correspondingly. The thickness of WTe$_2$ at each measurement location is later measured using atomic force microscopy. From these line-scans, we choose the regions where the probe is far away from the edge of WTe$_2$ so that the magnetic properties are uniform, to obtain spatial averages of $H_{r,1}$ and $\Delta A$, which are denoted $\overline{H}_{r,1,YIG/hBN}$ and $\overline{\Delta A}_{YIG/hBN}$ in the YIG/hBN region, and $\overline{H}_{r,1,YIG/WTe_2/hBN}$ and $\overline{\Delta A}_{YIG/WTe_2/hBN}$ in the YIG/WTe$_2$/hBN region, respectively. We further extract the $n = 1$ LM resonance field difference between two regions using $\Delta H_{r,1} = \overline{H}_{r,1,YIG/hBN} - \overline{H}_{r,1,YIG/WTe_2/hBN}$, as well as the Gilbert damping difference using $\Delta \alpha = \alpha_{YIG} \times \left( \sqrt{\overline{\Delta A}_{YIG/hBN} / \overline{\Delta A}_{YIG/WTe_2/hBN}} - 1 \right)$ as a function of the WTe$_2$ thickness. We note that the hBN overlayer does not change the Gilbert damping in YIG. The summarized results containing the data from both sample 1 and sample 2 are shown in Figs. 4a and 4b. Raw data from sample 2 can be found in Supporting Information Fig. S2. The thinnest WTe$_2$ acquired in the experiment is 3.2nm from sample 2, which is approximately the thickness of a quadruple-layer WTe$_2$.

Figures 4a and 4b indicate that both $\Delta H_{r,1}$ and $\Delta \alpha$ have almost no WTe$_2$ thickness dependence. There is a small sample-to-sample variation possibly due to different YIG/WTe$_2$ interfacial quality. The change of $n = 1$ LM resonance field, $\Delta H_{r,1}$, is as large as ~38 Oe even when the WTe$_2$ thickness approaches the quadruple-layer thickness. This indicates that the modification of magnetic anisotropy is due to the YIG/WTe$_2$ interfacial interaction, with no bulk contribution. For the increase of Gilbert damping $\Delta \alpha$, no obvious thickness dependence is observed when comparing the data from the same sample. In sample 2, the Gilbert damping enhancement due to the quadruple-layer WTe$_2$ has almost the same value as the 50 nm thick WTe$_2$ flake, indicating that no thickness dependence of spin pumping can be resolved from our measurement. There are two possible interpretations of these results. First, if the spin current injected into $WTe_2$ is mainly relaxed due to spin relaxation in the bulk, then the experimental result is a demonstration of ultra-short spin diffusion length along the *c* axis[38], smaller or comparable to the thinnest $WTe_2$ flake (3.2 nm), employed in this experiment. It is much smaller than the 8nm spin diffusion length in the in-plane direction measured using inverse spin Hall effect [22]. Note that due to the change in mobility and the metal-insulator transition in few layer $WTe_2$ when its thickness reduces [42], the spin diffusion length approximated here could be inaccurate. Alternatively, it is possible that the spin relaxation is primarily due to the interfacial SOC induced by inversion symmetry breaking at the interface and in the $WTe_2$ crystal lattice. In this case, the Gilbert damping enhancement will have no $WTe_2$ thickness dependence.

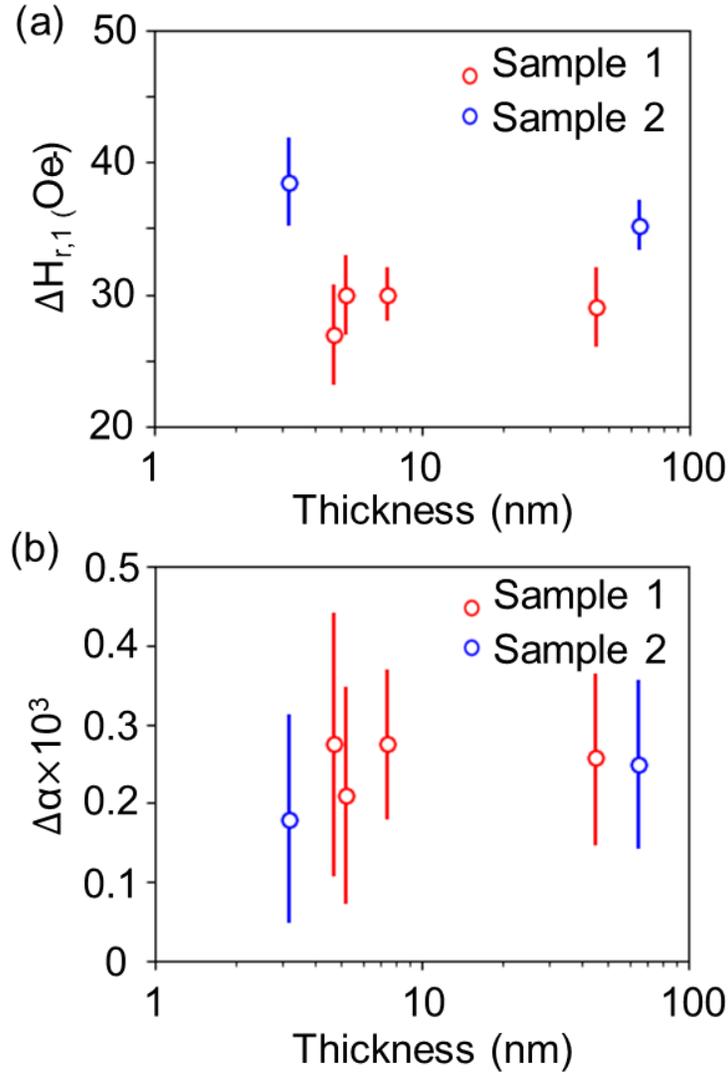

**Fig. 4 WTe₂ thickness dependence of resonance field and damping enhancement. a)** $H_{r,1}$ in the YIG/hBN and YIG/WTe$_2$/hBN regions are averaged respectively to get $\bar{H}_{r,1,\text{YIG}}$ and $\bar{H}_{r,1,\text{YIG/WTe}_2}$, and $\Delta H_{r,1} = \bar{H}_{r,1,\text{YIG}} - \bar{H}_{r,1,\text{YIG/WTe}_2}$. **b)** $\Delta\alpha$ as a function of WTe₂ thickness, and $\Delta\alpha = \alpha_{\text{YIG/WTe}_2} - \alpha_{\text{YIG}}$ where $\alpha_{\text{YIG}}$ is the Gilbert damping of bare YIG measured using broadband FMR for each sample, and

$$\alpha_{\text{YIG/WTe}_2} = \alpha_{\text{YIG}} \times \left(\sqrt{\overline{\Delta A}_{\text{YIG/hBN}} / \overline{\Delta A}_{\text{YIG/WTe}_2/\text{hBN}}}\right).$$

In conclusion, we have shown that the YIG/WTe₂ interface plays a critical role in both interfacial magnetic anisotropy and spin relaxation, making WTe₂ a promising material in magnetic memory applications. Combining the iPMA created by WTe$_2$ with the out-of-plane spin orbit torque generated by flowing a charge current along the *a* axis of WTe$_2$, one can possibly achieve field-free switching of a PMA magnetic cell for magnetic memory applications. It will improve the scalability, reduce the power consumption and increase operation speed of magnetic solid-state devices. Our result reveals new possibilities in selecting materials and designing spintronic devices. For example, one can consider other materials with low lattice symmetry and strong SOC to induce larger PMA type interfacial anisotropy in FMIs. To achieve a fully PMA material, one could utilize thinner FMIs to magnify the role of iPMA. Moreover, interfacial SOC also plays an important role in generating topologically protected magnetic textures in the FMIs [43]. These findings will motivate further research to reveal the fundamental physics arising at the interface between FMIs and nonmagnetic materials.

**Data availability:**

The data generated by the present study are available from the corresponding author on request.

**Supporting Information:**

A description of raw data on WTe$_2$ thickness dependence, a FMRFM measurement on a second sample, a FMRFM measurement at different RF frequency, a description of polarized Raman measurement result, and a detailed illustration of impact of broken mirror reflection symmetries on the magnetic anisotropy.

**Acknowledgements:**

This work was primarily supported by the Center for Emergent Materials: an NSF MRSEC under award number DMR-2011876 (GW, NV, YC, SG, FY, MR and PCH). KW and TT acknowledge support from the Elemental Strategy Initiative conducted by the MEXT, Japan (Grant Number JPMXP0112101001) and JSPS KAKENHI (Grant Numbers 19H05790, 20H00354 and 21H05233). DW, GC, CNL, and MB are supported by NSF under award DMR-2004801. We gratefully acknowledge N. Trivedi for insightful discussions. Fabrication and some characterization were performed in the Ohio State University NanoSystems Laboratory.

## Methods:

### Sample fabrication

Our YIG/WTe$_2$/hBN heterostructure was prepared by means of dry transfer and stacking [44]. hBN crystals were mechanically exfoliated under ambient conditions onto SiO$_2$/Si substrates (285 nm thick SiO$_2$). 20-40 nm thick hBN flakes were identified under an optical microscope and used for the capping layer for the stack. The hBN was picked up using a polymer-based dry transfer technique and then moved into an Ar-filled glove box with oxygen and water level below 0.1 ppm. Flux-grown WTe$_2$ crystals [45] were exfoliated inside the glove box and flakes with different thicknesses were optically identified and quickly picked up with the capping hBN layer then transferred to the YIG substrate. Finally, we removed the fully encapsulated sample from the glove box and performed the e-beam lithography and metallization (Ni/Au) step for alignment in our ferromagnetic resonance force microscope (FMRFM).

### Polarized Raman measurement

Polarized Raman spectra from the WTe$_2$ sample were collected using 633 nm excitation wavelength in an inVia Renishaw Raman microscope. The sample was loaded onto the microscope stage and positioned in such a way that the long edge of the flake was aligned parallel to the laser polarization (θ = 0°). In this configuration, the incident illumination is polarized vertically coming out of the laser and is aligned with the long axis of the WTe$_2$ flake. The polarization of the incident laser was rotated from 0 to 360° by 10° increments using a polarization rotator, while an analyzer was set to only allow vertically polarized light to enter the spectrometer. Raman spectra were collected at each polarization for 3 acquisitions with a 20 s time per acquisition. The laser power was set to 0.5 mW at the sample to avoid any damage by heating. Following spectral collection, the (baseline corrected) integrated intensities under each peak were calculated to make the contour plots and polar plots in Fig. 2c and 2d.

### FMRFM measurement and signal fitting

Our FMRFM performs locally measures FMR at room temperature in vacuum. The cantilever has natural frequency of ~18 KHz, spring constant of 0.2 N/m and Q factor of ~20000, resulting in force detection sensitivity of $10^{-15}$ N/Hz$^{1/2}$. The SmCo$_5$ magnetic particle attached on the cantilever has a magnetic moment of ~4 nemu. When a LM is on resonance, the local reduction of magnetization in out-of-plane direction will generate a stray field, which will couple the altered magnetization to the magnetic tip thus changing the cantilever oscillation amplitude and frequency. The change in cantilever oscillation is detected optically by laser fiber interferometry. Different LMs have different mode radii. For this microscopy study, we focus on $n = 1$ LM since it gives the highest spatial resolution. The $n = 1$ LM resonance peak is fit to a Lorentzian line shape, from which the peak position and peak height are extracted, which correspond to the resonance field $H_{r,1}$ and signal amplitude $\Delta A$.

# Enhancing Perpendicular Magnetic Anisotropy in Garnet Ferrimagnet by Interfacing with Few-Layer WTe$_2$

Guanzhong Wu[1*], Dongying Wang[1], Nishchhal Verma[1], Rahul Rao[2], Yang Cheng[1], Side Guo[1],

Guixin Cao[1], Kenji Watanabe[3], Takashi Taniguchi[4], Chun Ning Lau[1], Fengyuan Yang[1],

Mohit Randeria[1], Marc Bockrath[1], and P. Chris Hammel[1]

1. Department of Physics, The Ohio State University, Columbus, OH, 43210, USA

2. Materials and Manufacturing Directorate, Air Force Research Laboratory, Wright-Patterson Air Force Base, Dayton, OH, 45433, USA

3. Research Center for Functional Materials, National Institute for Materials Science, 1-1 Namiki, Tsukuba 305-0044, Japan

4. International Center for Materials Nanoarchitectonics, National Institute for Materials Science, 1-1 Namiki, Tsukuba 305-0044, Japan

*wu.2314@osu.edu

**FMRFM line-scan across edge of WTe₂ with different thickness**

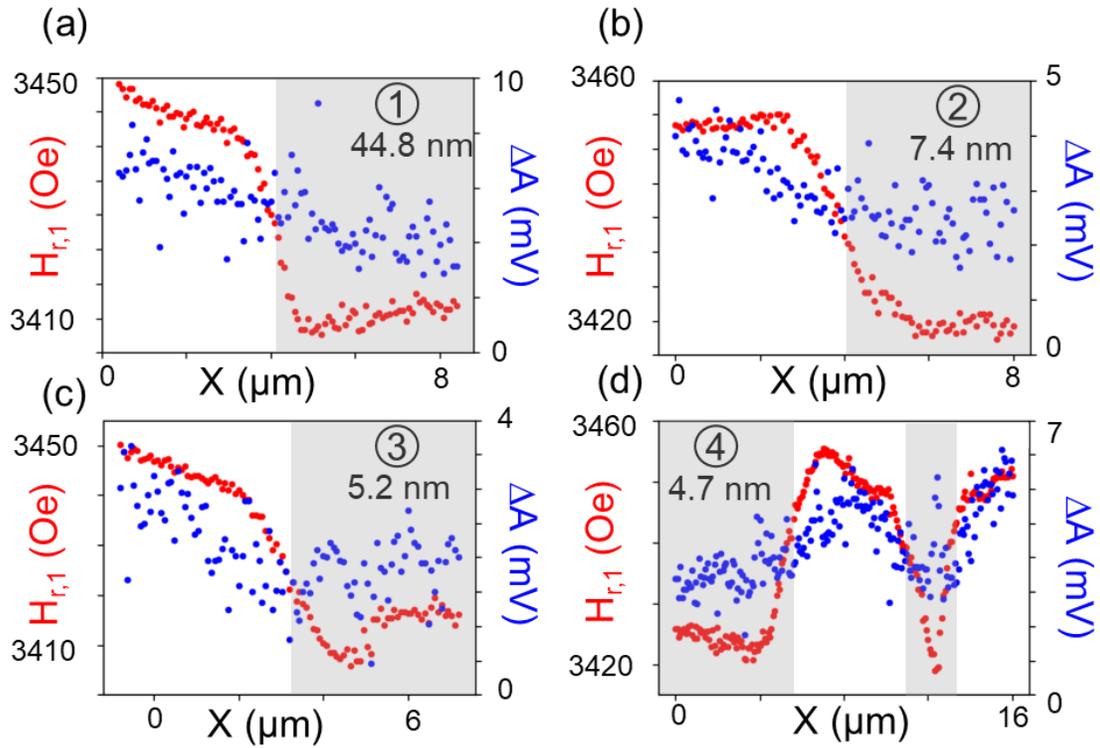

**Fig. S1 a-d,** FMRFM line-scans along the traces 1~4 indicated in Fig. 3b respectively. The gray shaded area in four figures are outlining the location of WTe$_2$ flake. $H_{r,1}$ and $\Delta A$ at each position are derived by fitting the $n=1$ LM to a Lorentzian line shape. The thickness of WTe$_2$ flake at each location are measured by atomic force microscope.

**FMRFM measurement on Sample 2**

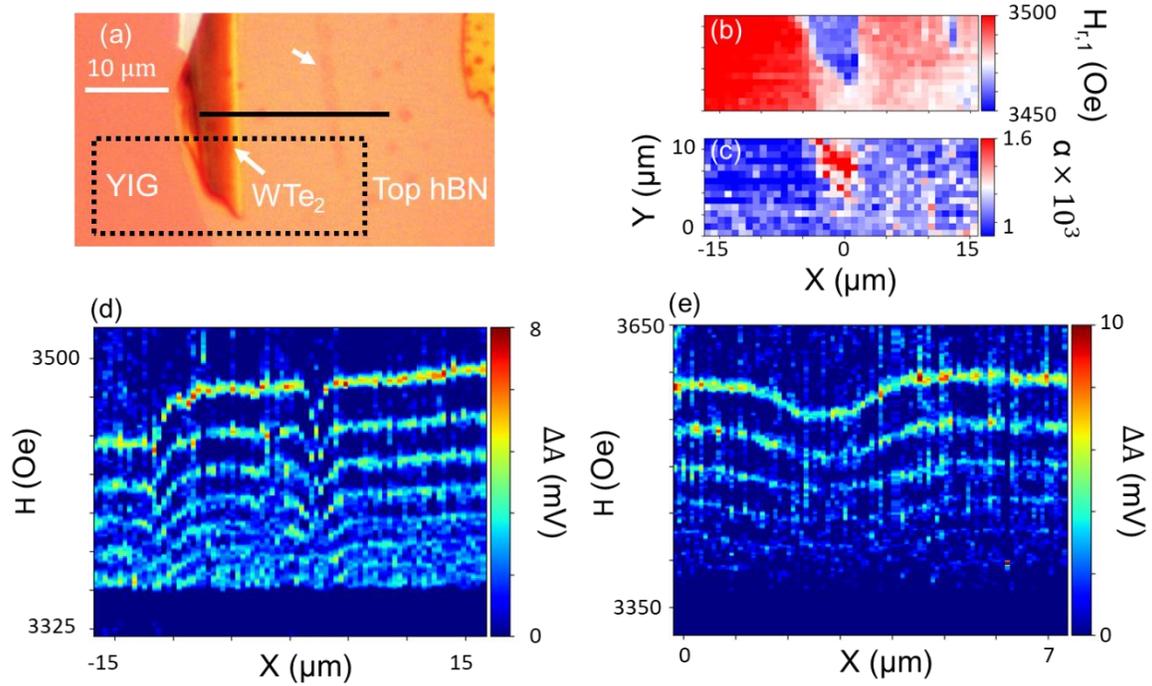

**Fig. S2 FMRFM measurement on sample 2. a,** The optical picture of the YIG/WTe$_2$/hBN heterostructure. **b,** 2D mapping of the $n = 1$ LM resonance field in the black dash line circled area. **c,** 2D mapping of the Gilbert damping extracted from $n = 1$ LM resonance peak amplitude in the black dash line circled area. **d,** FMRFM line scan along the trace indicated by the solid black line in Fig. S2a. A constant background is subtracted to show only the signal from the LMs resonance. **e,** Fine scan zoomed in on the quadruple layer WTe$_2$ stripe area

**FMRFMR measurement at 4 GHz**

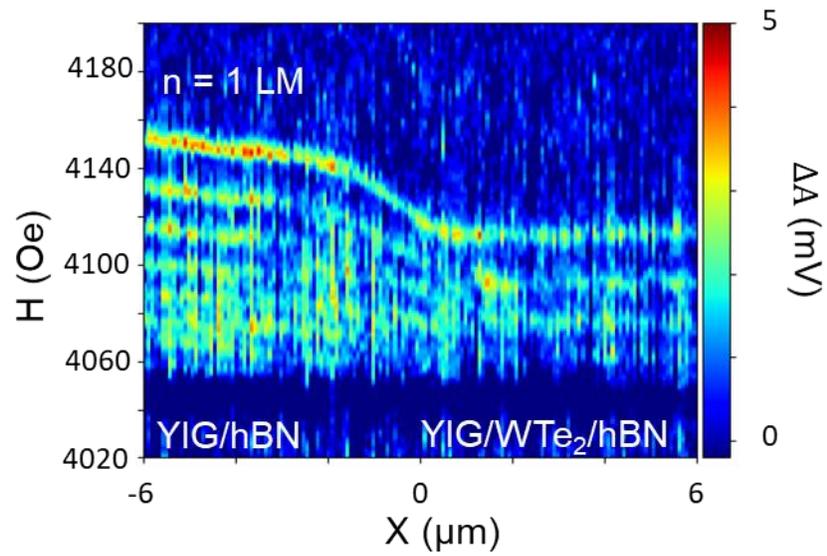

**Fig. S3 FMRFM measurement across WTe₂ edge at 4 GHz.** FMRFM line scan is measured at 4 GHz across the WTe₂ flake edge. The shift of the resonance field $H_{r,1}$ is 36 Oe, similar to the $H_{r,1}$ shift measured at 2 GHz. This result excludes the possibility that the resonance field shift arises from modification of the gyromagnetic ratio.

**Polarized Raman measurement**

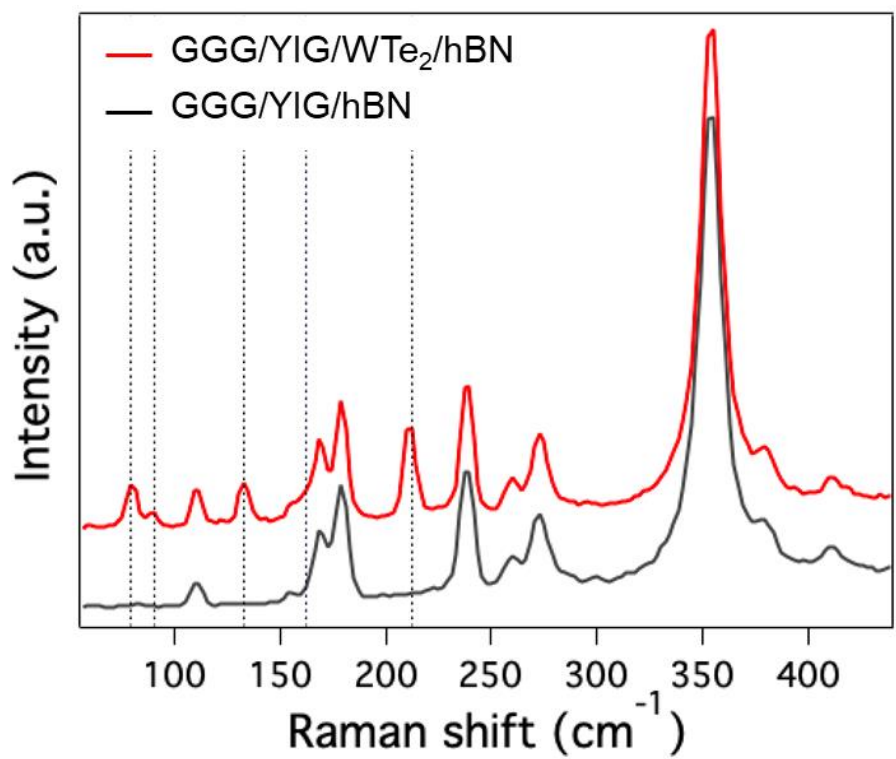

**Fig. S4 Polarized Raman measurement.** As shown by the red curve, the Raman spectrum taken on GGG/YIG/WTe$_2$/hBN heterostructure contains more peaks than WTe$_2$. The Raman spectrum taken in the GGG/YIG/hBN area identifies the peaks arising from the substrate GGG/YIG or top hBN encapsulation layer. By subtracting the Raman spectrum in the GGG/YIG/hBN area, the Raman spectra from WTe$_2$ layer are extracted and plotted in Fig. 2d. The black dash line are the markers indicating the Raman peaks of WTe$_2$

**Impact of broken mirror reflection symmetries on the magnetic anisotropy**

We describe theoretical constraints on the interface-induced magnetic anisotropy in the WTe$_2$/YIG bilayer. We first show that symmetry arguments alone do not provide strong constraints on the anisotropy tensor, given that we are dealing with an interface between two crystalline materials at an arbitrary orientation with respect to each other. We then present qualitative arguments, based on the interfacial spin-orbit coupling, that give insight into the magnetic anisotropy in WTe$_2$/YIG. This helps us understand why the *easy-axis* anisotropy that we observe in WTe$_2$/YIG differs from the results of Lee et al. [1] on YIG interfaces with a dozen different metallic and semiconducting materials, all of which exhibit interface-induced *easy-plane anisotropy* as predicted by theory [2].

On general grounds, the anisotropy (free) energy can be written as

$$\mathcal{F}_{anis} = \sum_{a,b} K_{ab}\, m_a\, m_b, \qquad \text{(S1)}$$

where *a* and *b* take on values *x,y,z*. We focus here on the leading term, quadratic in the magnetization, and ignore higher order anisotropy terms like $(m_x^4 + m_y^4 + m_z^4)$ or $(m_x^2 m_y^2 + m_y^2 m_z^2 + m_z^2 m_x^2)$. The form of $K_{ab} = K_{ba}$ is constrained by symmetry. Let us consider three cases, going from the most symmetric to the least.

Case I: The only broken symmetry is interfacial inversion (z → - z), which is relevant for the experiments of Ref. [1]. The magnetization is an axial vector (or pseudovector) that transforms under rotations like a vector but is unchanged under inversion. Thus $(m_x, m_y, m_z) \to (m_x, -m_y, -m_z)$ under reflection in a mirror plane with normal $\hat{x}$. Using reflection symmetry in mirror planes normal to $\hat{x}$ and to $\hat{y}$, we can see that all off-diagonal components of $K_{ab}$ vanish. Further, four-fold rotational symmetry about the $\hat{z}$ axis shows that $K_{xx} = K_{yy}$. Using $m_x^2 + m_y^2 + m_z^2 = 1$, we write $K_{xx}(m_x^2 + m_y^2)$ in terms of $m_z^2$, and defining $K_u = (K_{xx} - K_{zz})$, we obtain

$$\mathcal{F}_{anis} = -K_u\, m_z^2. \qquad (S2)$$

This symmetry analysis only constrains the *form* of the anisotropy energy, but not the *sign* of $K_u$. We will give below a simple microscopic argument [2] that shows that $K_u < 0$ (easy plane) for Case I.

Case II: In addition to broken interfacial inversion (z → - z), let us also break reflection symmetry in the plane normal to $\hat{x}$. This would be the case if the crystalline axes of WTe$_2$ were aligned with YIG. This also breaks four-fold rotational symmetry about $\hat{z}$, so that $K_{xx} \neq K_{yy}$. However, we can still use reflection symmetry in the plane normal to $\hat{y}$ to conclude that $K_{xy} = K_{yz} = 0$. Thus we find that

$$\mathbf{K} = \begin{pmatrix} K_{xx} & 0 & K_{xz} \\ 0 & K_{yy} & 0 \\ K_{xz} & 0 & K_{zz} \end{pmatrix} \qquad (S3)$$

Case III: When the crystalline axes of WTe$_2$ are *not* aligned with YIG, which is the experimentally relevant case, all mirror reflection and rotation symmetries are broken. Then there are no symmetry constraints on $K_{ab}$ and all six components of this symmetric tensor are in general non-zero.

Let us now see how, despite the lack of general symmetry-based constraints, we can still get some qualitative insight about the form of the anisotropy from simple microscopic considerations informed by symmetry. YIG is a ferrimagnetic Mott insulator, with two inequivalent Fe-sites coupled via antiferromagnetic (AFM) superexchange interactions. We thus focus on how interfacial spin-orbit coupling (SOC) impacts AFM superexchange.

The broken symmetry at the interface leads to an electric field $\mathcal{E} = -\nabla V(r)$, whose direction will be discussed in detail below for three cases. This in turn produces a magnetic field in the rest frame of the electron which underlies SOC. As the electron moves along $\hat{\mathbf{r}}_{ij}$ from site $\mathbf{i}$ to $\mathbf{j}$, it experiences an SOC field in the direction $\hat{\mathbf{d}}_{ij}$ which is determined by $\mathcal{E} \times \hat{\mathbf{r}}_{ij}$. The SOC Hamiltonian is thus given by $-i\lambda \sum_{\alpha\beta} c_{i\alpha}^{\dagger}(\hat{\mathbf{d}}_{ij} \cdot \boldsymbol{\sigma}_{\alpha\beta}) c_{j\beta}$. Including the effect of this term in addition to the usual hopping $t$ and Hubbard $U$ in the standard strong coupling expansion calculation leads to the Hamiltonian

$$\mathcal{H}_{ex} = J \sum_{<i,j>} \mathbf{S}_i \cdot \mathbf{S}_j + D \sum_{<i,j>} \hat{\mathbf{d}}_{ij} \cdot \mathbf{S}_i \times \mathbf{S}_j + K_0 \sum_{<i,j>} (\hat{\mathbf{d}}_{ij} \cdot \mathbf{S}_i)(\hat{\mathbf{d}}_{ij} \cdot \mathbf{S}_j). \tag{S4}$$

Here the spin $\mathbf{S}_i$ at site $\mathbf{i}$ is coupled to its neighbors via the AFM superexchange $J \sim \frac{t^2}{U}$ and the Dzyaloshinskii-Moriya interaction (DMI) $D \sim \frac{t\lambda}{U}$. The $K_0$ term will be the focus of our attention below as it leads to magnetic anisotropy. We note that the general form of $\mathcal{H}_{ex}$ is in fact substantially independent [2] of the microscopic mechanism and very similar results are obtained not only for superexchange but also for Zener double exchange and RKKY interactions.

Case I: Let us again return to the simplest case with broken interfacial inversion (z → - z). This leads to an electric field $\mathcal{E} = -\nabla V(r)$ along $\hat{z}$, the normal to the interface. The SOC magnetic field direction is then given by $\hat{\mathbf{d}}_{ij} = \hat{z} \times \hat{\mathbf{r}}_{ij}$; see Figure S4(a). This is the well-known Rashba SOC at interfaces. We note in passing that $\hat{\mathbf{d}}_{ij}$ is antisymmetric under the interchange of i and j, and thus leads to a DMI term where $\mathbf{S}_i \times \mathbf{S}_j$ is also antisymmetric.

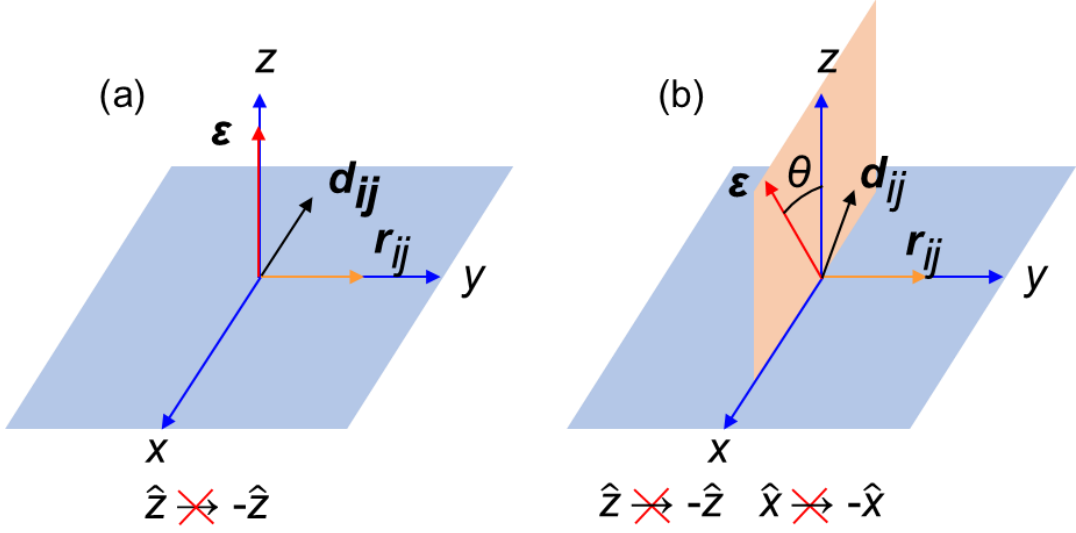

**Fig. S5 Symmetry based selection of magneto-crystalline anisotropy.** Interfacial SOC originates from an effective Electric field $\mathcal{E} = -\nabla V(r)$ whose direction is determined by the broken mirror planes in the system. This electric field leads to spin-orbit coupling (SOC), with the $\hat{\mathbf{d}}_{ij} = \hat{\mathcal{E}} \times \hat{\mathbf{r}}_{ij}$, the direction of the SOC magnetic field. Note that the direction of the electron hop $\hat{\mathbf{r}}_{ij}$ lies in the xy plane of the interface. As shown in the text $\hat{\mathbf{d}}_{ij}$ controls the interface-induced magnetic anisotropy. (a) When only surface inversion is broken, $\mathbf{d}_{ij}$ is constrained to lie in the interface and interfacial SOC leads to easy-plane anisotropy. (b) If there are other broken mirror planes, the $\mathbf{d}_{ij}$ must lie outside the interfacial plane. This can lead to a perpendicular magnetic anisotropy in systems like YIG/WTe$_2$ bilayers.

We see that in Case I, $\hat{\mathbf{d}}_{ij}$ lies *in the plane* of the interface, and the third term in eq. (S4) then takes the form $K_0 \sum_{\mathbf{i}}(S_{\mathbf{i}}^x S_{\mathbf{i}+y}^x + S_{\mathbf{i}}^y S_{\mathbf{i}+x}^y)$ for a square lattice. To make the connection with magnetic anisotropy, we look at a continuum approximation with a slowly varying magnetization $\mathbf{m}(\mathbf{r})$. We make a Taylor expansion of $\mathbf{S}_r$ in terms of its value at $r$, denoted by $\mathbf{m}(\mathbf{r})$, and its spatial derivatives. The exchange and DMI terms involve gradients of $\mathbf{m}(\mathbf{r})$, but we focus here on local terms that do not involve derivatives to understand the magnetic anisotropy. The leading term is $+ K_0(m_x^2 + m_y^2)$ which can be rewritten as $- K_0 m_z^2$ using the fact that $m_x^2 + m_y^2 + m_z^2 = 1$ at each $\boldsymbol{r}$. Thus, we may identify $K_0$ with the anisotropy $K_u$ defined in eq. (S2).

The microscopic analysis leads to the result $K_0 = \frac{-\lambda^2}{U} < 0$ and this explains the *easy-plane anisotropy* arising Rashba SOC at the interface. The easy-plane nature of the anisotropy is in fact a general feature of various microscopic models as emphasized in Ref. [2]. We note however that these authors used the opposite sign convention for anisotropies from the one we use here. The easy plane vs. easy-axis character is, of course, independent of sign conventions. The FMR experiments of Ref. [1] have seen the interface-induced *easy-plane anisotropy* predicted by the theory in a YIG interfaces with several metallic and semiconducting materials.

The key difference between the YIG/WTe2 bilayer studied here and systems studied earlier [1] is that WTe2 has a broken mirror plane (the *ac* plane) as shown in Fig. 1(a) of the paper. We now look at the effect of this lower symmetry on the microscopic analysis.

Case II: Let us break reflection symmetry in the plane normal to $\hat{x}$ in addition to broken interfacial inversion. We choose $\hat{x}$ parallel to the *b* axis, $\hat{y}$ parallel to *a,* and $\hat{z}$ parallel to *c*. Reflection symmetry in the $\hat{y}$ mirror plane constrains the electric field $\mathcal{E} = -\boldsymbol{\nabla} V(\boldsymbol{r})$ to lie in the xz plane, at an angle $\theta$ from the z-axis as shown in Fig. S5(b). Thus

$$\mathbf{d}_{ij} = (\sin\theta\,\hat{x} + \cos\theta\,\hat{z}) \times \hat{\mathbf{r}}_{ij} \tag{S5}$$

where $\hat{r}_{ij}$ is a vector in the interface (*xy* plane) and $0 \leq \theta \leq \pi$. Using eq. (S5), we may rewrite the last term in the Hamiltonian (S4) as

$$K_0 \sin^2\theta \sum_i \left(S_i^z S_{i+y}^z\right) + K_0 \cos^2\theta \sum_i \left(S_i^x S_{i+y}^x + S_i^y S_{i+x}^y\right)$$

$$- K_0 \sin\theta \cos\theta \sum_i \left(S_i^z S_{i+y}^x + S_i^x S_{i+y}^z\right)$$

As before, we make a continuum approximation with a smoothly varying $\mathbf{m}(\mathbf{r})$ and focus only on the local terms, without gradients, to obtain the magnetic anisotropy. We find that the leading order contribution to anisotropy is $-K_0 \cos 2\theta\, m_z^2 + K_0 \sin 2\theta\, m_z m_x$. This analysis correctly captures the non-zero $K_{xz}$ expected on general grounds; see eq. (S3). We did not include here, for simplicity, the effects of broken four-fold rotation that would have led to $K_{xx} \neq K_{yy}$.

Case III: When we lose all mirror symmetries, the case relevant to the YIG/WTe$_2$ experiment, the electric field $\mathcal{E} = -\nabla V(\mathbf{r})$ will point in a general direction specified by $0 \leq \theta \leq \pi$ and $0 \leq \varphi \leq 2\pi$, and there will be no symmetry constraints on the anisotropy tensor $K_{ab}$.

Let us conclude by highlighting the key qualitative difference between Case I on the one hand and Cases II and III on the other. In Case I, the only broken symmetry is interfacial inversion ($z \rightarrow -z$). Then symmetry constrains the $\hat{\mathbf{d}}_{ij}$, the direction of the SOC **B**-field, to lie in the plane of the interface and this leads to easy-plane anisotropy as described above. In Cases II and III, there are other additional broken mirror planes, and this leads to the $\hat{\mathbf{d}}_{ij}$ vector being pulled out of the plane of the interface. This immediately leads to the possibility of an easy-axis like character to the anisotropy, although in the general case one has a non-trivial anisotropy tensor $K_{ab}$.